\documentclass{aa}

\usepackage{graphicx}
\usepackage[varg]{txfonts}
\usepackage{natbib}
\usepackage[pdftex,colorlinks=true,linkcolor=blue,citecolor=blue,urlcolor=blue]{hyperref}

\def\im{\mathrm{i}}

\def\expo#1{\mathbf{e}^{#1}}

\def\H{\mathcal{H}}
\def\G{\mathcal{G}}
\def\red#1{{\color{red}#1}}
\DeclareMathOperator{\lcm}{lcm}

\graphicspath{{./figures/}}

\newcommand\figI{
  \begin{figure*}
    \centering
    \includegraphics[width=\linewidth]{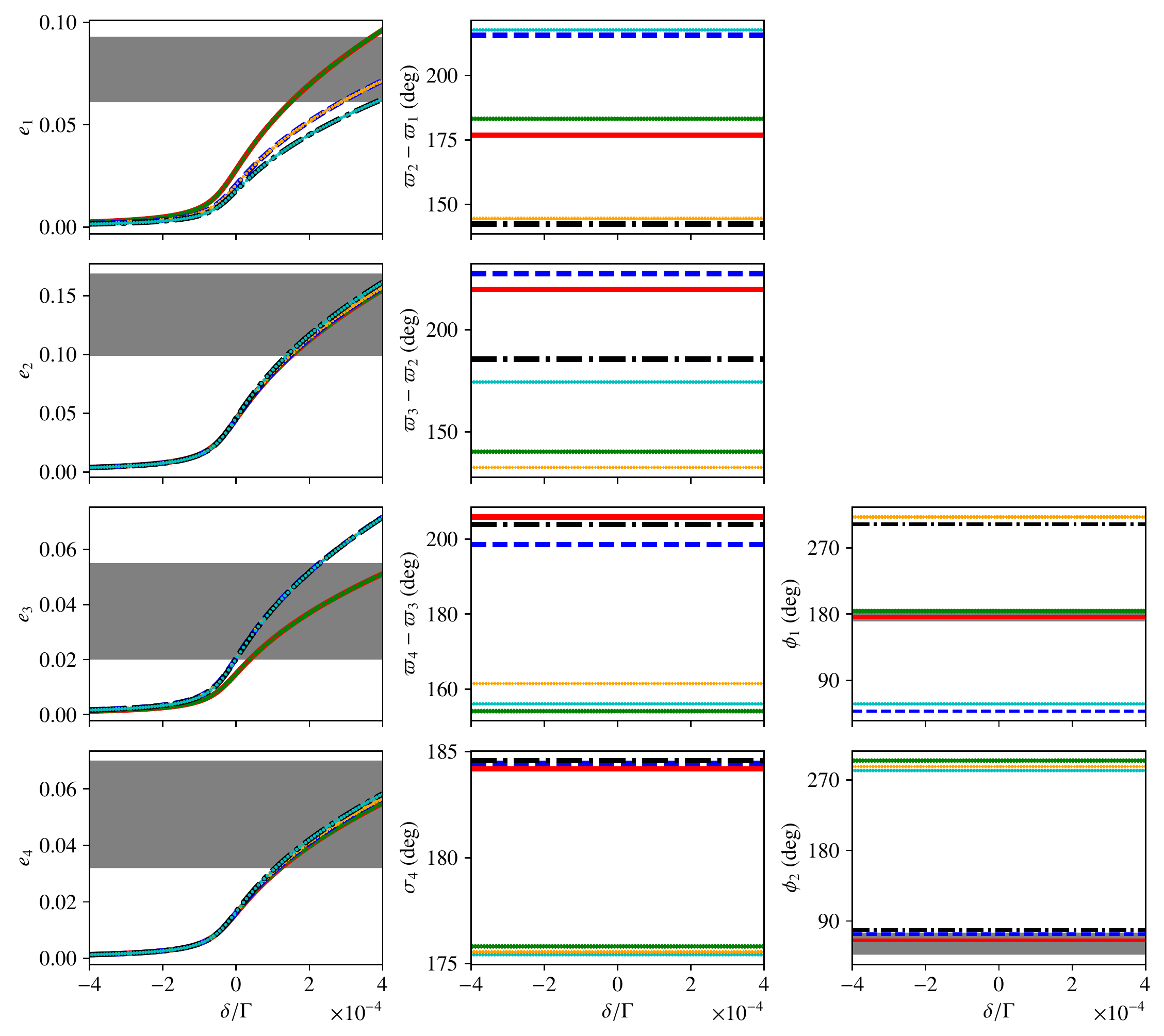}
    \caption{Location of the (six) fixed points
    for the \object{Kepler-223} resonant chain (4 planets),
    as a function of the parameter $\delta/\Gamma$ (see Sect.~\ref{sec:model}).
    I plot the eccentricities of the planet (\textit{left}),
    the difference of longitudes of periastron between two successive planets,
    as well as the two-planet resonant angle $\sigma_4$ (\textit{center}),
    and the Laplace angles (\textit{right}).
    Each fixed point is represented by the same color in all the plots.
    The gray bands (on the \textit{left} and \textit{right} columns)
    show the observed parameters of the planets \citep[taken from][]{mills_resonant_2016}.
    For the eccentricities (\textit{left}), the gray bands correspond to the $1-\sigma$ uncertainties.
    For the Laplace angles, they correspond to the observed libration amplitudes
    \citep[which are well constrained by TTVs, see][]{mills_resonant_2016}.
    The observed positions of the planets (and especially the Laplace angles),
    are consistent with a libration around the red fixed point (see also Fig.~\ref{fig:II}).
    }
    \label{fig:I}
  \end{figure*}
}

\newcommand\figII{
  \begin{figure}
    \centering
    \includegraphics[width=\linewidth]{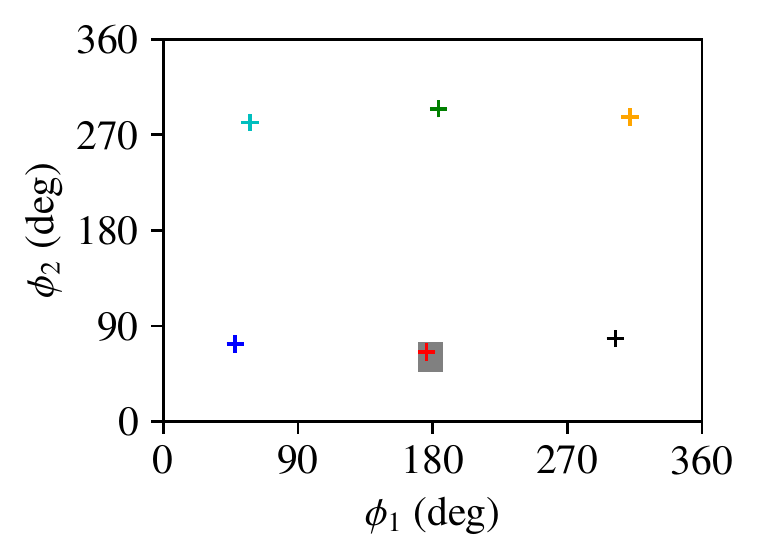}
    \caption{Comparison of the observed libration amplitudes of
    the Laplace angles in the \object{Kepler-223} system
    \citep[gray rectangle, values taken from][]{mills_resonant_2016},
    with the positions of the (six) fixed points as determined from the analytical model.
    I use the same colors as in Fig.~\ref{fig:I}.
    The observations are consistent with a libration around the red fixed point.}
    \label{fig:II}
  \end{figure}
}

\newcommand\figIII{
  \begin{figure}
    \centering
    \includegraphics[width=\linewidth]{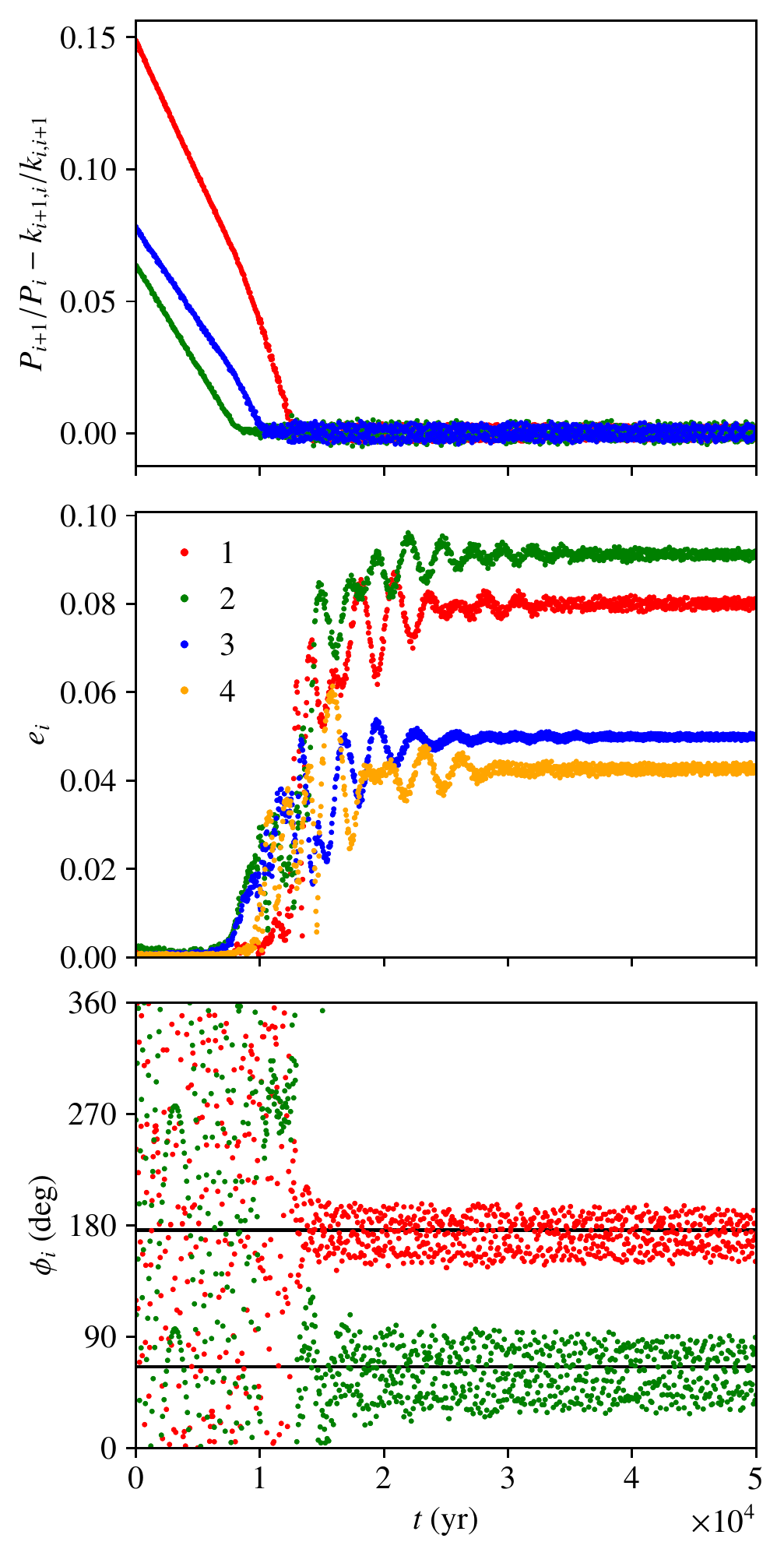}
    \caption{Example of a simulation that successfully reproduces the
    observed configuration of the \object{Kepler-223} system.
    The scenario of capture is of type BCA (see Sect.~\ref{sec:constraints}).
    I plot the period ratio between
    each consecutive pair of planets (\textit{top}),
    the planets' eccentricities (\textit{middle}),
    and the two Laplace angles (\textit{bottom}).
    The two horizontal black lines in the \textit{bottom} plot
    represent the equilibrium values of the Laplace angles expected from the
    analytical model (see Sect.~\ref{sec:equilibrium}).}
    \label{fig:III}
  \end{figure}
}

\newcommand\tabI{
  {\setlength{\tabcolsep}{0.2ex}
  \begin{table}
    \begin{center}
      \caption{Statistics of capture of Kepler-223
      around the six possible equilibrium configurations
      (see Figs.~\ref{fig:I} and \ref{fig:II}),
      as a function of the order in which the planets have been captured in the
      resonant chain.}
      \begin{tabular}{c|cccccc|c}
\hline
order     &                        ABC &                        ACB &                        BAC &                        BCA &                        CAB &                        CBA & mean\\
\#        &                       1543 &                        516 &                       1343 &                        980 &                        506 &                        870 & \\
\hline
\rule{0pt}{4mm}
\red{red} & \red{$ 4.1_{-0.5}^{+0.6}$} & \red{$23.1_{-1.9}^{+2.0}$} & \red{$10.9_{-0.9}^{+0.9}$} & \red{$52.0_{-1.6}^{+1.6}$} & \red{$26.1_{-2.0}^{+2.1}$} & \red{$35.1_{-1.7}^{+1.7}$} & \red{25.2}\\[1mm]
green     &      $32.0_{-1.2}^{+1.2}$  &      $14.5_{-1.6}^{+1.7}$  &      $52.3_{-1.4}^{+1.4}$  &      $11.5_{-1.0}^{+1.1}$  &      $ 9.1_{-1.3}^{+1.5}$  &      $22.9_{-1.4}^{+1.5}$  & 23.7\\[1mm]
blue      &      $ 3.3_{-0.5}^{+0.5}$  &      $17.6_{-1.7}^{+1.8}$  &      $ 4.8_{-0.6}^{+0.7}$  &      $18.5_{-1.3}^{+1.3}$  &      $28.1_{-2.0}^{+2.1}$  &      $31.1_{-1.6}^{+1.6}$  & 17.2\\[1mm]
orange    &      $39.0_{-1.3}^{+1.3}$  &      $18.4_{-1.7}^{+1.9}$  &      $ 5.7_{-0.6}^{+0.7}$  &      $ 4.3_{-0.6}^{+0.7}$  &      $11.9_{-1.5}^{+1.6}$  &      $ 1.7_{-0.4}^{+0.6}$  & 13.5\\[1mm]
black     &      $ 3.4_{-0.5}^{+0.5}$  &      $18.0_{-1.7}^{+1.9}$  &      $ 1.4_{-0.3}^{+0.4}$  &      $ 1.7_{-0.4}^{+0.5}$  &      $16.6_{-1.7}^{+1.8}$  &      $ 2.9_{-0.6}^{+0.7}$  &  7.3\\[1mm]
cyan      &      $18.3_{-1.0}^{+1.0}$  &      $ 8.3_{-1.2}^{+1.4}$  &      $24.9_{-1.2}^{+1.2}$  &      $11.9_{-1.0}^{+1.1}$  &      $ 8.3_{-1.2}^{+1.4}$  &      $ 6.3_{-0.8}^{+0.9}$  & 13.0\\[1mm]
\hline
      \end{tabular}
      \tablefoot{A stands for the capture of planets 1,2 in the 4/3 MMR,
      B for the capture of planets 2,3 in the 3/2 MMR,
      and C for the capture of planets 3,4 in the 4/3 MMR.
      I run 6000 N-body simulations varying initial conditions,
      5758 of which were captured in the 3:4:6:8 resonant chain.
      For each possible order of capture, I give the number of simulations
      that were captured in this order,
      and the statistics of captures
      (percentages and 1-$\sigma$ confidence interval)
      around each equilibrium.
      The mean values (\textit{right} column) are computed
      assuming an equal probability for each capture order.
      The system is currently observed around the red equilibrium
      (the corresponding row is highlighted in red).}
      \label{tab:I}
    \end{center}
  \end{table}}
}

\begin{document}

\title{Analytical model of multi-planetary resonant chains\\
  and constraints on migration scenarios}
\titlerunning{Analytical model of resonant chains}
\author{J.-B. Delisle\inst{1,2}}
\institute{Observatoire de l'Université de Genève, 51 chemin des Maillettes, 1290, Sauverny, Switzerland\\
  \email{jean-baptiste.delisle@unige.ch}
    \and ASD, IMCCE, Observatoire de Paris - PSL Research University, UPMC Univ. Paris 6, CNRS,\\
    77 Avenue Denfert-Rochereau, 75014 Paris, France
}

\date{\today}

\abstract{
Resonant chains are groups of planets for which
each pair is in resonance, with an orbital period ratio
locked at a rational value
(2/1, 3/2, etc.).
Such chains naturally form as a result of convergent migration
of the planets in the proto-planetary disk.
In this article, I present an analytical model of
resonant chains of any number of planets.
Using this model, I show that a system captured in a resonant chain
can librate around several possible equilibrium configurations.
The probability of capture around each equilibrium
depends on how the chain formed,
and especially on the order in which the planets have been captured
in the chain.
Therefore, for an observed resonant chain,
knowing around which equilibrium the chain is librating allows for constraints to be put on the formation and migration scenario of the system.
I apply this reasoning to the four planets orbiting \object{Kepler-223}
in a 3:4:6:8 resonant chain.
I show that the system is observed around one of the six equilibria predicted by
the analytical model.
Using N-body integrations, I show that
the most favorable scenario to reproduce the observed configuration is
to first capture the two intermediate planets, then the outermost, and
finally the innermost.
}


\maketitle


\section{Introduction}
\label{sec:introduction}

Mean motion resonances (MMR) between two planets are a natural outcome of
the convergent migration of planets in a gas-disk
\citep[e.g.,][]{weidenschilling_orbital_1985}.
The planets initially form farther away from each other,
and planet-disk interactions induce a migration of the planets.
The period ratio between the planets decreases
until they get captured in a MMR.
The planets then continue to migrate whilst maintaining
their period ratio at a rational value (2/1, 3/2, etc.).
The eccentricities increase due to the resonant interactions,
until they reach an equilibrium between the migration
torque and the eccentricity damping exerted by the disk.
The argument of the resonance,
which is a combination of the mean longitudes of the two planets,
enters into libration (oscillations around an equilibrium value).

For systems of three and more planets,
once a pair of planets has been captured in a MMR,
the other planets might also join
this couple to form a chain of resonances.
Each time a planet gets captured in the chain,
it enters into a MMR (and thus maintains a constant and rational period ratio)
with each of the other planets of the chain.
The eccentricities of the planets and the resonant arguments of each pair
find a new equilibrium.
Such multi-planetary resonant chains are expected
from simulations of planet migration
\citep[e.g.,][]{cresswell_evolution_2006}.
Recently, \citet{mills_resonant_2016} showed that
the four planets in the Kepler-223 system are in
a 3:4:6:8 resonant chain
(period ratios of 4/3, 3/2, and 4/3 between consecutive pairs of planets).
Using transit timing variations (TTVs), the authors observed
that the Laplace angles of the system are librating with small amplitudes.
The Laplace angles are combinations of the mean longitudes of three
planets in the chain, and the observation of their libration is
evidence that the system is indeed captured in the resonant chain.
Using numerical simulations,
\citet{mills_resonant_2016} showed that the observed orbital configuration
is very well reproduced by
a smooth convergent migration of the planets.

In this article, I present an analytical model of resonant chains.
Analytical models have already been proposed,
in particular to study the dynamics of the Laplace resonance (1:2:4 chain)
between Io, Europa, and Ganymede \citep[e.g.,][]{henrard_orbital_1983}.
However, while several numerical studies have been dedicated to the
capture of planets in various resonant chains
\citep[e.g.,][]{cresswell_evolution_2006,papaloizou_dynamics_2010,
libert_trapping_2011,papaloizou_consequences_2016},
general analytical models have not yet been proposed.
Recently, \citet{papaloizou_three_2015} proposed a semi-analytical
model of three-planet resonances taking into account
only the interactions between consecutive planets in the chain,
with a particular focus on the \object{Kepler-60} system
\citep[12:15:20 resonant chain, see also][]{steffen_transit_2012,gozdziewski_laplace_2016}.
This model is very similar to the studies of the Laplace resonance between
the Galilean moons,
but is not well suited in the general case.
For instance, four-planet (or more) resonances are not considered.
Moreover, for some three-planet resonances,
the interactions between non-consecutive planets cannot be neglected.
For instance, in a 3:4:6 resonant chain, each planet is locked
in a first-order resonance with each of the other planets.
In particular, the innermost and outermost planets are involved in a
2/1 MMR that strongly influences the dynamics of the system.
I describe here a general model of resonant chains,
with any number of planets, valid for any resonance order.
I particularly focus on finding the equilibrium configurations
(eccentricities, resonant arguments, etc.)
around which a resonant system should librate.
While a real system may be observed with significant
amplitude of libration around the equilibrium, or could even have
some angles circulating, the position of the equilibria
still provides useful insights into the dynamics of the system.
In Sect.~\ref{sec:model}, I describe this analytical model,
and the method I use to find the equilibrium configurations.
In Sect.~\ref{sec:application}, I apply the model to \object{Kepler-223}.
I show that six equilibrium configurations
exist for this resonant chain, and that the system is observed
to be librating around one of them.
I also show that knowing the current configuration
of the system allows for interesting constraints to be put on its migration scenario,
and in particular on the order in which the planets have been captured
in the chain.

\section{Model}
\label{sec:model}
I consider a planetary system with $n$ planets
(which I denote with indices $1,...,n$
from the innermost to the outermost)
orbiting around a star (index 0).
I assume that the system is coplanar
and is locked in a chain of resonances.
In such a resonant chain, each pair of planets is locked in a MMR.
For two planets $i<j$,
I denote by $k_{j,i}/k_{i,j}$ the resonant ratio,
such that
\begin{equation}
  k_{j,i} n_j - k_{i,j} n_i \approx 0,
\end{equation}
where $n_i$ ($n_j$) is the mean motion of planet $i$ ($j$).
I also introduce the degree of the resonance
between planet $i$ and planet $j$
\begin{equation}
  q_{i,j} = k_{j,i} - k_{i,j}.
\end{equation}
At low eccentricities,
resonances of a lower degree have a
stronger influence on the dynamics of the system.

In order to study the dynamics of these resonant chains,
I generalize to $n$ planets the method
developped in the case of two-planet resonances
\citep{delisle_dissipation_2012,delisle_resonance_2014}.
The Hamiltonian of the system takes the form \citep{laskar_analytical_1991}
\begin{equation}\label{eq:hamposvel}
  \H = -\sum_{i=1}^n \G\frac{m_0m_i}{2 a_i}
  + \sum_{1\leq i<j\leq n} \left(-\G\frac{m_i m_j}{||\vec{r}_i-\vec{r}_j||}
      + \frac{\vec{\tilde{r}}_i.\vec{\tilde{r}}_j}{m_0} \right),
\end{equation}
where
$\G$ is the gravitational constant,
$m_i$ is the mass of body $i$,
$a_i$ is the semi-major axis,
$\vec{r}_i$ the position vector,
and $\vec{\tilde{r}}_i$ the canonically conjugated momentum
of planet $i$ \citep[in astrocentric coordinates, see][]{laskar_analytical_1991}.
The first sum on the right-hand side of Eq.~(\ref{eq:hamposvel}) is the Keplerian part of the Hamiltonian
(planet-star interactions),
while the second sum is the perturbative part (planet-planet interactions).

In the coplanar case (which I assume here)
the system has $2 n$ degrees of freedom (DOF),
with 2 DOF (4 coordinates) associated to each planet.
As for two-planet resonances \citep[e.g.,][]{delisle_dissipation_2012},
the number of DOF can be reduced by using the conservation of the total
angular momentum (1 DOF), and by averaging over the fast angles (1 DOF).
Therefore, the problem can be reduced to $2(n-1)$ DOF.
Even with these reductions, the phase space is still very complex,
especially for systems of many planets such as
\object{Kepler-223} (chain of 4 planets, 6 DOF),
and the problem is, in most cases, non-integrable.
In this study, I focus on finding the fixed point of the averaged problem,
which provides useful insight into the dynamics of the system,
and especially into the values around which the angles of a resonant system should librate.
The method described in the following is a generalization of the method presented in
\citet{delisle_dissipation_2012} which focuses on finding the fixed points
for two-planet MMR.

I denote by $\lambda_i$ and $\varpi_i$ the mean longitude and longitude of periastron of planet $i$ (in astrocentric coordinates), respectively.
The actions canonically conjugated to the angles
$\lambda_i$ and $-\varpi_i$ are the circular angular momentum $\Lambda_i$
and the angular momentum deficit \citep[AMD, see][]{laskar_spacing_2000}
$D_i$, respectively.
These actions are defined as follows
\begin{eqnarray}
  \Lambda_i &=& \beta_i\sqrt{\mu_i a_i},\\
  D_i &=& \Lambda_i-G_i = \Lambda_i \left(1-\sqrt{1-e_i^2}\right),
\end{eqnarray}
where
$G_i=\Lambda_i\sqrt{1-e_i^2}$ is the angular momentum of planet $i$,
$\beta_i = m_i m_0/(m_0+m_i)$,
$\mu_i = \G (m_0+m_i)$.
At low eccentricities the deficit of angular momentum $D_i$
is proportional to $e_i^2$.
The Hamiltonian (Eq.~(\ref{eq:hamposvel})) can be expressed
using these action-angle coordinates
\begin{equation}\label{eq:hamLaD}
  \H = - \sum_{i=1}^n \frac{\mu_i^2\beta_i^3}{2\Lambda_i^2}
  + \sum_{1\leq i<j\leq n} \H_{i,j}(\Lambda_i,\Lambda_j,D_i,D_j,\lambda_i,\lambda_j,\varpi_i,\varpi_j),
\end{equation}
where the first sum is the Keplerian part,
which depends only on $\Lambda_i$ (or equivalently $a_i$),
and $\H_{i,j}$ is the perturbation between the planets $i$ and $j$,
which depends on the eight coordinates associated to $i$ and $j$.
I follow the method described in \citet{laskar_stability_1995}
to compute $H_{i,j}$ as a power series of the eccentricities
(or equivalently of $\sqrt{D_i}$ and $\sqrt{D_j}$),
and a Fourrier series of the angles,
where the coefficients are functions of
$\Lambda_i$ and $\Lambda_j$ (i.e., of the semi-major axes).
For a system that is close to the resonance or resonant,
the semi-major axes remain close to the nominal resonant values (Kepler's third law)
\begin{equation}\label{eq:nominalsma}
  \frac{a_i}{a_j} \approx \frac{a_{i,0}}{a_{j,0}} = \left(\frac{k_{i,j}}{k_{j,i}}\right)^{2/3} \left(\frac{\mu_i}{\mu_j}\right)^{1/3}.
\end{equation}
I introduce
\begin{equation}
  \Delta\Lambda_i = \Lambda_i - \Lambda_{i,0},
\end{equation}
where
\begin{equation}
  \Lambda_{i,0} = \beta_i\sqrt{\mu_i a_{i,0}},
\end{equation}
and expand the Keplerian part at degree 2,
and the perturbative part at degree 0 in $\Delta\Lambda_i$
\begin{eqnarray}\label{eq:hamdLaD}
  \H &=& \sum_{i=1}^n n_{i,0} \Delta\Lambda_i
  - \frac{3}{2}\frac{n_{i,0}}{\Lambda_{i,0}}\Delta\Lambda_i^2\nonumber\\
  &+& \sum_{1\leq i<j\leq n} \H_{i,j}(D_i,D_j,\lambda_i,\lambda_j,\varpi_i,\varpi_j),
\end{eqnarray}
where $n_{i,0}$ is the nominal mean motion of planet $i$, such that
\begin{equation}
 \frac{n_{i,0}}{n_{j,0}} = \frac{k_{j,i}}{k_{i,j}}.
\end{equation}
The perturbative part does not depend on $\Lambda_i$ anymore,
but is simply evaluated at $\Lambda_{i,0}$.

In order to perform the reductions associated
to the conservation of angular momentum and to the averaging,
I first change the system of coordinates.
For the sake of readability,
I present the general case (with any number of planets, in any resonance of any degree)
in Appendix~\ref{sec:generalchgcoord},
and take here the example
of a system of four planets in a 3:4:6:8 resonant chain
(as is the case for \object{Kepler-223}).
I introduce new canonically conjugated angles and actions
as follows (see Eqs.~(\ref{eq:angles}) and (\ref{eq:actions}))
\begin{eqnarray}\label{eq:k223angacts}
  \phi_1    = \lambda_1 + \lambda_3 - 2 \lambda_2,
  &\qquad&  L_1    = \Lambda_1,\nonumber\\
  \phi_2    = \lambda_2 + 2\lambda_4 - 3 \lambda_3,
  &\qquad&  L_2    = 2 \Lambda_1 + \Lambda_2,\nonumber\\
  \phi_3    = \lambda_3 - \lambda_4,
  &\qquad&  \Gamma = 8 \Lambda_1 + 6\Lambda_2 + 4\Lambda_3+ 3\Lambda_4,\nonumber\\
  \phi_4    = 4 \lambda_4 - 3 \lambda_3,
  &\qquad&  G = G_1 + G_2 + G_3 + G_4,\nonumber\\
  \sigma_1  = 4 \lambda_4 - 3 \lambda_3 - \varpi_1,
  &\qquad&  D_1,\nonumber\\
  \sigma_2  = 4 \lambda_4 - 3 \lambda_3 - \varpi_2,
  &\qquad&  D_2,\nonumber\\
  \sigma_3  = 4 \lambda_4 - 3 \lambda_3 - \varpi_3,
  &\qquad&  D_3,\nonumber\\
  \sigma_4  = 4 \lambda_4 - 3 \lambda_3 - \varpi_4,
  &\qquad&  D_4.
\end{eqnarray}
$G$ is the total angular momentum, and is a conserved quantity;
Its canonically conjugated angle ($\phi_4$) does not appear in the Hamiltonian.
The angle $\phi_3$ is the only fast angle, and the averaging of the Hamiltonian is done over this angle.
Therefore, its conjugated action ($\Gamma$) is constant in the average problem.
The averaging is simply done by discarding all terms that depends on $\phi_3$
in the Fourrier expansion of the Hamiltonian.%
\footnote{I restrict this study to first order in the planet-star mass ratio,
which means that three-planet terms (of order two in the mass) are neglected.}
The angle $\phi_1$ is the argument of the Laplace resonance between the three innermost planets.
The angle $\phi_2$ is the argument of the Laplace resonance between the three outermost planets.

The two outer planets (3 and 4) may seem to play an important
role in Eq.~(\ref{eq:k223angacts}), but this is only due to the arbitrary choice
of canonical coordinates (many other choices are possible).
Any two-planet resonant angle can be expressed as a combination
of the angles of Eq.~(\ref{eq:k223angacts}).
The arguments of the 4/3 resonance between the two outermost planets
are $\sigma_3$ and $\sigma_4$.
The arguments of the 3/2 resonance between planets 2 and 3
are
$\sigma_2 - 2 \phi_2 = 3\lambda_3 - 2\lambda_2 - \varpi_2$ and
$\sigma_3 - 2 \phi_2 = 3\lambda_3 - 2\lambda_2 - \varpi_3$.
The arguments of the 4/3 resonance between planets 1 and 2
are
$\sigma_1 - 2 \phi_2 - 3 \phi_1 = 4\lambda_2 - 3\lambda_1 - \varpi_1$ and
$\sigma_2 - 2 \phi_2 - 3 \phi_1 = 4\lambda_2 - 3\lambda_1 - \varpi_2$.
Arguments of resonances between non-consecutive pairs can also
be expressed in the same way.
For instance, for the 2/1 resonance between planets 1 and 3,
the arguments are
$\sigma_1 - 2 \phi_2 - \phi_1 = 2\lambda_3 - \lambda_1 - \varpi_1$ and
$\sigma_3 - 2 \phi_2 - \phi_1 = 2\lambda_3 - \lambda_1 - \varpi_3$.
For a system of $n$ planets captured in a resonant chain,
$\phi_i$ ($i\leq n-2$) and $\sigma_i$ ($i\leq n$)
(and all their linear combinations) librate around equilibrium values.
All the actions also oscillate around equilibria.
These equilibrium values correspond to stable fixed points of the average problem.

For this example, I expand the perturbative part
at first order in eccentricities ($\sqrt{D_i}$),
and obtain an expression of the form
\begin{eqnarray}\label{eq:hamdLaDk223}
  \H &=& -\frac{3}{2}\sum_{i=1}^3 \frac{n_{i,0}}{\Lambda_{i,0}}\Delta\Lambda_i^2\nonumber\\
  &+& C_{1,2} \sqrt{D_1} \cos(\sigma_1 - 2 \phi_2 - 3 \phi_1)\nonumber\\
  &+& C_{2,1} \sqrt{D_2} \cos(\sigma_2 - 2 \phi_2 - 3 \phi_1)\nonumber\\
  &+& C_{1,3} \sqrt{D_1} \cos(\sigma_1 - 2 \phi_2 - \phi_1)\nonumber\\
  &+& C_{3,1} \sqrt{D_3} \cos(\sigma_3 - 2 \phi_2 - \phi_1)\nonumber\\
  &+& C_{2,3} \sqrt{D_2} \cos(\sigma_2-2\phi_2)
  + C_{3,2} \sqrt{D_3} \cos(\sigma_3-2\phi_2)\nonumber\\
  &+& C_{2,4} \sqrt{D_2} \cos(\sigma_2-\phi_2)
  + C_{4,2} \sqrt{D_4} \cos(\sigma_4-\phi_2)\nonumber\\
  &+& C_{3,4} \sqrt{D_3} \cos(\sigma_3)
  + C_{4,3} \sqrt{D_4} \cos(\sigma_4),
\end{eqnarray}
where the first term of Eq.~(\ref{eq:hamdLaD}) vanishes (because $\Gamma$ is constant),
and $C_{i,j}$ are constant coefficients that depend on
the masses and nominal semi-major axes ($a_{i,0}$).
I provide explicit formulas for the case of
the 3:4:6:8 resonant chain in Appendix~\ref{sec:coef346}.
Since the Hamiltonian (Eq.~(\ref{eq:hamdLaDk223}))
is developed at first order in eccentricities,
only first-order resonances appear.
In particular, the 3/8 resonance between planets 1 and 4
is neglected since it would only appear at order 5 in eccentricities.
In order to use a consistent (canonical) set of coordinates,
$\Delta \Lambda_i$ should be replaced in Eq.~(\ref{eq:hamdLaDk223}) by
\begin{eqnarray}
  \Delta\Lambda_1 &=& \Delta L_1,\nonumber\\
  \Delta\Lambda_2 &=& \Delta L_2 - 2\Delta L_1,\nonumber\\
  \Delta\Lambda_3 &=& \Delta L_1 - 3(\Delta L_2+\epsilon),\nonumber\\
  \Delta\Lambda_3 &=& 2 \Delta L_2 + 4\epsilon,
\end{eqnarray}
where
\begin{equation}
  \epsilon = D - \delta = \sum_{i=1}^n \Delta\Lambda_{i},
\end{equation}
which measures the distance of the system to the exact resonance,
where\begin{equation}
  D = \sum_{i=1}^n D_i,
\end{equation}
is the total deficit of angular momentum,
and
\begin{equation}
 \delta = \sum_{i=1}^n \Lambda_{i,0} - G
\end{equation}
is the nominal total deficit of angular momentum (at exact resonance).
Since $\Lambda_{i,0}$ and $G$ are constants, $\delta$ is also
a conserved quantity and can be used as a parameter instead of $G$.
The other parameter $\Gamma$ does not appear explicitly in the Hamiltonian,
but is hidden in the values of $\Lambda_{i,0}$, $a_{i,0}$, and $n_{i,0}$.
Indeed, the nominal semi-major axis ratios are fixed at the resonant values
(Eq.~(\ref{eq:nominalsma})),
but $\Gamma$ sets the global scale of the system
($\Gamma = 8 \Lambda_{1,0} + 6\Lambda_{2,0} + 4\Lambda_{3,0} + 3\Lambda_{4,0}$, for a 3:4:6:8 chain).
The value of $\Gamma$ does not influence the dynamics
of the system appart from changing the
scales of distance, time, and energy \citep{delisle_dissipation_2012}.
Therefore, one only need to vary the value of $\delta/\Gamma$ to
study the evolution of the phase space (in particular the positions of fixed points).

For a given value of $\delta/\Gamma$, the fixed points
are found by solving the following system of equations
\begin{eqnarray}\label{eq:Hamjaceqs}
  \dot{\sigma}_i &=& \frac{\partial\H}{\partial D_i} = 0 \qquad (i\leq n),\nonumber\\
  \dot{D}_i &=& -\frac{\partial\H}{\partial \sigma_i} = 0 \qquad (i\leq n),\nonumber\\
  \dot{\phi}_i &=& \frac{\partial\H}{\partial L_i} = 0 \qquad (i\leq n-2),\nonumber\\
  \dot{L}_i &=& -\frac{\partial\H}{\partial \phi_i} = 0 \qquad (i\leq n-2).
\end{eqnarray}
This is a system of $4(n-1)$ equations, with $4(n-1)$ unknowns ($2(n-1)$ DOF),\
which in general possesses a finite number of solutions.
These solutions can correspond to elliptical (stable) fixed points or hyperbolic (unstable) ones.
To assess the stability of fixed points, I compute the eigenvalues of
the linearized equations of motions around the fixed point.
Stable fixed points have purely imaginary eigenvalues,
while the eigenvalues around unstable fixed points have a non-zero real part.

\section{Application to Kepler-223}
\label{sec:application}

In this section, I apply my model to the four planets
orbiting in a 3:4:6:8 resonant chain around \object{Kepler-223}.
In Sect.~\ref{sec:equilibrium},
I focus on the comparison between the positions of the stable fixed points
and the observed configuration of the system
(values of resonant angles, eccentricities).
In Sect.~\ref{sec:constraints}, I derive constraints on the
order in which the planets have been captured in the resonant chain,
from the observation of the equilibrium around which the system is currently librating.

\subsection{Equilibrium configurations}
\label{sec:equilibrium}

\figI

\figII

I follow the method described in Sect.~\ref{sec:model},
and expand the Hamiltonian at first order in eccentricities (Eq.~(\ref{eq:hamdLaDk223})),
to solve for the positions of the fixed points (Eq.~(\ref{eq:Hamjaceqs}))
as a function of the parameter $\delta/\Gamma$
(see Appendix \ref{sec:fixedpoints} for more details).
I only consider here the stable (elliptical)
fixed points that correspond to the libration in resonance.
In the case of \object{Kepler-223},
I find six families of stable fixed points (parameterized by $\delta/\Gamma$),
corresponding to six possible areas of libration for the system.
I show in Fig.~\ref{fig:I} the positions of these fixed points,
and compare them with the observed values
\citep[taken from][]{mills_resonant_2016}.
As shown by \citet{mills_resonant_2016}, the TTVs constrain the values
of $\phi_1$ and $\phi_2$ very well, while the eccentricities are only roughly determined.
The two-planet resonant angles $\sigma_i$
(which depend on the longitudes of periastron)
are not well constrained \citep{mills_resonant_2016},
so the theoretical values cannot be compared to the observations.
As $\delta/\Gamma$ increases, the eccentricities increase,
but the angles ($\phi_i$, $\sigma_i$) remain constant.
Since the best observational constraints are on the values of
$\phi_1$ and $\phi_2$,
I compare in Fig.~\ref{fig:II} the observed values of these angles,
with the six possible equilibrium values (which are independent of $\delta/\Gamma$).
The observations correspond very well to
a libration of the system around one of the six configurations (the red one).
This result confirms that the system is captured in the resonant chain,
and that my model is correct in first approximation.
It is interesting to wonder why the system was captured
around this particular configuration and not one of the five others.
This might simply be by chance (probability of 1/6 for each configuration),
but might also greatly depend on the migration scenario.
Observing around which equilibrium the resonant chain is librating
could provide interesting constraints on the migration undergone by the planets.

\subsection{Constraints on migration scenarios}
\label{sec:constraints}

\tabI

\figIII

As explained in Sect.~\ref{sec:introduction},
when two planets are captured in a MMR, their eccentricities
increase until they reach an equilibrium \citep[e.g.,][]{delisle_dissipation_2012},
their period ratio remains locked at the resonant value,
and the arguments of the resonance librate.
For a resonance between two planets, and at low eccentricities,
the equilibrium configuration is unique.
When additional planets join the chain,
the system evolves toward a new equilibrium (eccentricities, two-planet resonant angles, Laplace angles).
However, when more than two planets are involved in the chain,
there can be more than one equilibrium
(see Sect.~\ref{sec:equilibrium}),
and the probability of capture around each of the new equilibria
is not necessarily equal.
Moreover, this probability might depend on the order in which the planets
are captured in the chain.
Indeed, the planets that are already captured in resonance
have their eccentricities excited, and the angles associated
to the already formed resonances are librating around an equilibrium,
while planets that are not yet captured should have lower eccentricities,
and their mean longitudes should be randomly distributed.
Therefore, the initial conditions (angles, eccentricities) at the
moment of the capture of a planet in the chain
greatly depend on which planets are already in the chain.

In this section, I investigate how the order in which the planets are
captured in the resonant chain influences the probability
of capture around each of the six equilibria found
in the case of \object{Kepler-223} (see Sec.~\ref{sec:equilibrium}).
This problem is very complex (the phase space has 6 DOF).
In particular, each time an additional planet joins the chain,
the system might cross one or several separatrices
before being captured around one of the new equilibria.
I do not attempt to treat this problem analytically, but
rather numerically estimate the probabilities of capture
by running 6000 N-body simulations
including prescriptions for the migration,
with varying initial conditions.
I use the same integrator as in \citet{delisle_stability_2015},
and set constant timescales for the migration torque and the eccentricity
damping of each planet.
The migration timescales ($T_{\text{mig.},i}$) are set to
$10^{6}$, $3\times 10^{5}$, $2\times 10^{5}$, and $1.5\times 10^{5}$~yr
(from the innermost to the outermost planet).
The eccentricity damping timescales ($T_{\text{ecc.},i}$) are set
such that for each planet $T_{\text{mig.},i}/T_{\text{ecc.},i} = 50$.
The planets start with circular and coplanar orbits.
The innermost planet is at 1~AU, and $\lambda_1 = 0$,
and the other planets' semi-major axes and mean longitudes
are randomly drawn.
The mean longitudes are drawn from a uniform distribution between 0 and $2\pi$.
The semi-major axis ratio $a_2/a_1$ is uniformly drawn in the range $[1.23, 1.3]$, $a_3/a_2$ in the range $[1.32, 1.38]$,
and $a_4/a_3$ in the range $[1.23, 1.28]$.
This allows the planet pairs to be captured in MMR in any possible order.
The orbits are integrated for $5\times 10^4$~yr.

Among the 6000 simulations, 5758 are captured in the 3:4:6:8 resonant chain,
while 242 are captured in other chains (or not captured).
All of the 5758 captured simulations are observed to librate around one
of the six equilibria predicted by my analytical model.
This confirms that this first-order model is correctly describing the dynamics of
the resonant chain.
For each simulation, I check in which order the planets were captured,
and around which equilibrium the system ended.

The results are shown in Table~\ref{tab:I}.
I use `A' to denote the capture of the two innermost planets (1,2)
in the 4/3 MMR,
`B' the capture of the intermediate ones (2,3) in the 3/2 MMR,
and `C' the capture of the outermost planets (3,4) in the 4/3 MMR.
Thus, ABC means that the planets were captured in the chain from the innermost
to the outermost one, and so on.
For each possible order of capture (ordering of A, B, and C),
I compute the percentage of capture around each equilibrium.
If the captures were equally probable,
the percentages would all be of about $16.7\%$ (1/6).
This is clearly not the case for \object{Kepler-223}.
For instance, the black equilibrium is difficult to reach (low probability)
except in the cases ACB and CAB (see Table~\ref{tab:I}).
This means that the two innermost planets and the two outermost ones
must first be captured in two independent two-planet resonances,
and then the two pairs join to form the four-planet resonant chain.

Since the system is currently observed around the red equilibrium
(see Figs.~\ref{fig:I} and \ref{fig:II}),
it is interesting to look at the corresponding capture probabilities.
The most favorable case is BCA,
with a probability of $52\%$ to capture the system around the observed equilibrium
(Table~\ref{tab:I}).
This case corresponds to a capture of planets 2 and 3 in the 3/2 MMR,
then planet 4 joins the chain (2:3:4),
and finally planet 1 is captured to form the 3:4:6:8 chain.
Figure~\ref{fig:III} shows an example of a simulation
following this scenario, and reproducing the observed
configuration of the \object{Kepler-223} system.
Conversely, in the case ABC
(capture from the innermost to the outermost planet),
the probability to reproduce the observed
configuration is only $4.1\%$.
The mean capture probability around the observed configuration
(assuming equal probability for each capture order),
is the highest value ($25.2\%$).
It is thus not surprising to observe the system in this configuration,
rather than in one of the five others.

\section{Discussion}
\label{sec:discussion}

In this article I describe an analytical model of resonant chains.
The model is valid for any number of planets involved in the chain,
and for any resonances (of any order).
In particular, I use it to
determine the equilibrium configurations around which
a resonant chain librates.
I show that contrarily to two-planet MMR,
multiple equilibria may exist, even at low eccentricities,
when three or more planets are involved.

I specifically study the case of
the four planets around \object{Kepler-223} which
have been confirmed to be captured in a 3:4:6:8 resonant chain
\citep[using TTVs, see][]{mills_resonant_2016}.
Using the analytical model expanded at first order in eccentricities,
I show that six equilibrium configurations exist for this system,
and the planets might have been captured around any of
these six equilibria.
However, the capture probabilities are not the same for each equilibrium,
and depend on the order in which the planets have been captured in the chain.
Using N-body integrations including migration prescriptions,
I show that the scenario the most capable of reproducing the observed configuration
of the system is to first capture the intermediate planets (2 and 3) in the 3/2 MMR,
then to capture the outermost planet (4) to form a 2:3:4 chain between the
three outermost planets,
and finally to capture the innermost planet (1).
This scenario of capture reproduces the observed configuration in
$52\%$ of the simulations,
while capturing the planets from the innermost to the outermost
reproduces the observed configuration for only $4.1\%$ of the simulations.
It should be noted that several hypotheses are made to compute these statistics.
The planets are initially outside the resonances, with period ratios slightly
higher than the resonant values.
The migration and eccentricity damping timescales are fixed for each planet,
such that the period ratio between each pair decreases (convergent migration).
The planets are initially on circular and coplanar orbits.
I only vary the initial semi-major axes
and initial mean longitudes of the planets.
The statistics would probably slightly change with
a different set-up for the simulations,
but this would not change the two main results of this study:
\begin{enumerate}
\item Resonant chains can be captured around
  several equilibrium configurations (six for \object{Kepler-223}),
\item observing a system around one of the possible equilibria provides useful constraints on the scenario of formation and migration of the planets.
\end{enumerate}
Other properties of the resonant chain might provide
useful additional constraints on such a scenario.
For instance, in the case of \object{Kepler-223},
the two inner planets, as well as the two outer ones,
are in a compact 4/3 resonance.
If these planets formed with much wider separations,
they must have avoided permanent capture into the 2/1 and 3/2 resonances
to reach the currently observed 4/3 resonance.
However, determining a complete scenario
for the formation of the \object{Kepler-223} system
is a highly degenerated problem
and is beyond the scope of this article.

\begin{acknowledgements}
  I thank the anonymous referee for his/her useful comments.
  I acknowledge financial support from the Swiss National Science Foundation (SNSF).
  This work has, in part, been carried out within the framework of
  the National Centre for Competence in Research PlanetS
  supported by SNSF.
\end{acknowledgements}

\bibliographystyle{aa}
\bibliography{laplace}

\appendix

\section{Change of coordinates in the general case}
\label{sec:generalchgcoord}

In this Appendix I describe
the change of coordinates corresponding to Eq.~(\ref{eq:k223angacts})
in the general case.
I introduce the angles
\begin{eqnarray}\label{eq:angles}
  \phi_i &=& \frac{1}{c_i} \left(
  \frac{k_{i,i+1}}{q_{i,i+1}}\lambda_i
  +\frac{k_{i+2,i+1}}{q_{i+1,i+2}}\lambda_{i+2}
  \right.\nonumber\\
  &&\phantom{\frac{1}{c_i} (}
  \left.
  -\left(\frac{k_{i+1,i}}{q_{i,i+1}}+\frac{k_{i+1,i+2}}{q_{i+1,i+2}} \right)\lambda_{i+1}
  \right) \qquad (i \leq n-2),\nonumber\\
  \phi_{n-1} &=& \lambda_{n-1} - \lambda_n,\nonumber\\
  \phi_{n} &=& \frac{k_{n,n-1}\lambda_n
    - k_{n-1,n}\lambda_{n-1}}{q_{n-1,n}},\nonumber\\
  \sigma_i &=& \phi_n - \varpi_i,
\end{eqnarray}
where the coefficients $c_i$ are renormalizing factors given by Eq.~(\ref{eq:ci}).
The actions canonically conjugated to $\sigma_i$ are the
AMD of the planets ($D_i$),
while the actions canonically conjugated to $\phi_i$ are
\begin{eqnarray}\label{eq:actions}
  L_i &=& \sum_{j=1}^i c_i l_{i,j} \Lambda_j \qquad (i\leq n-2),\nonumber\\
  \Gamma &=& \frac{k_{n-1,n}}{q_{n-1,n}}\left(
    \Lambda_n + \frac{k_{n,n-1}}{k_{n-1,n}} \left(
    \Lambda_{n-1} + \frac{k_{n-1,n-2}}{k_{n-2,n-1}} \left(
    ... + \frac{k_{2,1}}{k_{1,2}}\Lambda_1 \right)\right)\right),\nonumber\\
  G &=& \sum_{i=1}^n \Lambda_i - D_i = \sum_{i=1}^n G_i,
\end{eqnarray}
where
\begin{equation}\label{eq:lij}
  l_{i,j} = \frac{1}{k_{i,i+1}}
    \sum_{r=j}^i q_{r,r+1} \prod_{s=r+1}^{i}\frac{k_{s+1,s}}{k_{s-1,s}}
    \qquad (j\leq i \leq n-2).
\end{equation}

It can be shown that the Hamiltonian does not depend on the angle $\phi_n$,
and the conjugated action $G$ (total angular momentum of the system)
is thus conserved (as expected).
Since $\phi_{n-1}$ is the only fast angle in the new system of coordinates,
the averaging is done over this angle,
and $\Gamma$ (conjugate of $\phi_{n-1}$)
is thus constant in the averaged problem.
The remaining $2(n-1)$ DOF (once $G$ and $\Gamma$ are fixed),
are thus defined by the angles $\sigma_i$ ($i\leq n$) and $\phi_i$ ($i\leq n-2$),
and their conjugated actions $D_i$ ($i\leq n$) and $L_i$ ($i\leq n-2$).
The angles $\phi_i$ ($i\leq n-2$) are the Laplace resonant angles
between each group of three consecutive planets.
The angles $\sigma_{n-1}$ and $\sigma_{n}$ are the classical
two-planet resonant angles between the two outermost planets,
and any two-planet (not necessarily consecutive)
resonant angle can be expressed as a combination
of the angles $\sigma_i$ ($i\leq n$) and $\phi_i$ ($i\leq n-2$).

In order to express the Hamiltonian in these new coordinates,
one first needs to invert the change of coordinates
(i.e., express $\Lambda_i$, $\lambda_i$, etc. as functions of $L_i$, $\phi_i$, etc.).
For the angles, the inverse transformation reads
\begin{eqnarray}\label{eq:invangles}
  \varpi_i &=& \phi_n - \sigma_i,\nonumber\\
  \lambda_i &=& \phi_n
    + \frac{k_{n-1,n}}{q_{n-1,n}}\prod_{j=i}^{n-1} \frac{k_{j+1,j}}{k_{j,j+1}} \phi_{n-1}
    + \sum_{j=i}^{n-2} c_j l_{j,i} \phi_j.
\end{eqnarray}
Then, the Fourier expansion of the perturbative part of the averaged Hamiltonian exhibits angles of the form
\begin{equation}
  d (k_{j,i} \lambda_j - k_{i,j} \lambda_i) + p \varpi_i + (d q_{i,j} - p) \varpi_j
\end{equation}
Since all these angles should be expressed as combinations of $\sigma_i$ ($i \leq n$) and $\phi_i$ ($i\leq n-2$),
one have to make sure that only integers appear in the combinations.
Indeed, if one of the coefficients is a fraction, then the corresponding angle will not be $2\pi$-periodic.
For instance, if one of the angles appearing in the Fourrier series is $3/2 \phi_1 + ...$, then
$\phi_1 = 0$ is not equivalent to $\phi_1 = 2\pi$.
One should also make the coefficients as small as possible,
to avoid introducing an unnecessary periodicity in the Hamiltonian
(and thus an artificial duplication of fixed points).
I thus choose the renormalizing factors $c_i$ to obtain the smallest possible integers in the
Fourier expansion of the perturbative part.
\begin{equation}\label{eq:ci}
  c_i = \frac{\lcm_{r\leq i, r<s}(D_{i,r,s})}{\gcd_{r\leq i, r<s}(N_{i,r,s})}
  \qquad (i \leq n-2),
\end{equation}
where
\begin{equation}
  \frac{N_{i,r,s}}{D_{i,r,s}} = k_{s,r} l_{i,s} - k_{r,s} l_{i,r}
  \qquad (l_{i,s} = 0 \text{ for } s>i)
\end{equation}
is the fraction reduced to its simplest form.
The Hamiltonian is expanded in power series of the eccentricities,
and truncated at a given degree $d_\text{max}$.
Only the combinations for which the degree $q_{r,s} = k_{s,r}-k_{r,s} \leq d_\text{max}$
appear in the truncated Hamiltonian.
Therefore, only these combinations should be considered in the computation
of the coefficient $c_i$ in Eq.~(\ref{eq:ci}).

For the actions, the inverse change of coordinates reads
\begin{eqnarray}\label{eq:invacts}
  \Lambda_i &=& \frac{k_{i,i+1}}{c_i q_{i,i+1}} L_i
    + \frac{1}{c_{i-2}}\frac{k_{i,i-1}}{q_{i-1,i}} L_{i-2}\nonumber\\
    && - \frac{1}{c_{i-1}}\left(\frac{k_{i,i-1}}{q_{i-1,i}} + \frac{k_{i,i+1}}{q_{i,i+1}}\right) L_{i-1}
    \qquad (i\leq n-2),\nonumber\\
  \Lambda_{n-1} &=& \Gamma - \frac{k_{n-1,n}}{q_{n-1,n}} (G+D)
    + \frac{1}{c_{n-3}}\frac{k_{n-1,n-2}}{q_{n-2,n-1}} L_{n-3}\nonumber\\
    && - \frac{1}{c_{n-2}}\left(\frac{k_{n-1,n-2}}{q_{n-2,n-1}} + \frac{k_{n-1,n}}{q_{n-1,n}}\right) L_{n-2},\nonumber\\
  \Lambda_n &=& \frac{k_{n,n-1}}{q_{n-1,n}} (G+D)- \Gamma
    + \frac{1}{c_{n-2}}\frac{k_{n,n-1}}{q_{n-1,n}} L_{n-2},
\end{eqnarray}
where
\begin{equation}
  D = \sum_{i=1}^n D_i = \sum_{i=1}^n \Lambda_i - G
\end{equation}
is the total deficit of angular momentum.
I introduce
\begin{equation}
 \delta = \sum_{i=1}^n \Lambda_{i,0} - G,
\end{equation}
which is the nominal total deficit of angular momentum (at exact resonance).
Since $\Lambda_{i,0}$ and $G$ are constants, $\delta$ is also
a conserved quantity.
I additionally define
\begin{equation}
  \epsilon = D - \delta = \sum_{i=1}^n \Delta\Lambda_{i},
\end{equation}
which measures the distance of the system to the exact resonance,
such that
\begin{eqnarray}\label{eq:invactionsbis}
  \Delta\Lambda_i &=& \frac{k_{i,i+1}}{c_i q_{i,i+1}} \Delta L_i
    + \frac{1}{c_{i-2}}\frac{k_{i,i-1}}{q_{i-1,i}} \Delta L_{i-2}\nonumber\\
    && - \frac{1}{c_{i-1}}\left(\frac{k_{i,i-1}}{q_{i-1,i}} + \frac{k_{i,i+1}}{q_{i,i+1}}\right) \Delta L_{i-1}
    \qquad (i\leq n-2),\nonumber\\
  \Delta \Lambda_{n-1} &=& - \frac{k_{n-1,n}}{q_{n-1,n}} \epsilon
    + \frac{1}{c_{n-3}}\frac{k_{n-1,n-2}}{q_{n-2,n-1}} \Delta L_{n-3}\nonumber\\
    && - \frac{1}{c_{n-2}}\left(\frac{k_{n-1,n-2}}{q_{n-2,n-1}} + \frac{k_{n-1,n}}{q_{n-1,n}}\right) \Delta L_{n-2},\nonumber\\
  \Delta \Lambda_n &=& \frac{k_{n,n-1}}{q_{n-1,n}} \epsilon
    + \frac{1}{c_{n-2}}\frac{k_{n,n-1}}{q_{n-1,n}} \Delta L_{n-2},
\end{eqnarray}
with $\Delta L_i = L_i - \sum_{j\leq i} c_i l_{i,j}\Lambda_{j,0}$.

\section{Coefficients of the Hamiltonian for a 3:4:6:8 resonant chain}
\label{sec:coef346}

In this appendix, I provide the expressions of the coefficients
$C_{i,j}$ that appear in the Hamiltonian of a 3:4:6:8 resonant chain
at first order in eccentricities (see Eq.~\ref{eq:hamdLaDk223}).
I follow the method described in \citet{laskar_stability_1995}.
Each coefficient is a sum of two components,
coming from the direct and indirect part of the perturbation
\begin{eqnarray}
  C_{i,j} &=& -\G \frac{m_i m_j}{a_{j,0}} \sqrt{\frac{2}{\Lambda_{i,0}}}\left( C_{i,j}^\text{dir}
  - \frac{\G m_0}{\sqrt{\mu_i\mu_j\alpha_{i,j}}} C_{i,j}^\text{ind}
  \right),\nonumber\\
  C_{j,i} &=& -\G \frac{m_i m_j}{a_{j,0}} \sqrt{\frac{2}{\Lambda_{j,0}}}\left( C_{j,i}^\text{dir}
  - \frac{\G m_0}{\sqrt{\mu_i\mu_j\alpha_{i,j}}} C_{j,i}^\text{ind}
  \right),
\end{eqnarray}
where $i<j$, $\alpha_{i,j} = a_{i,0}/a_{j,0}$,
and
\begin{eqnarray}
C_{i,j}^\text{dir} &=& \frac{2}{3} \alpha_{i,j}^{-1} b_{3/2}^{(1)}
  - b_{3/2}^{(0)} + \frac{7}{6} \alpha_{i,j} b_{3/2}^{(1)}
  - \frac{5}{2} \alpha_{i,j}^{2} b_{3/2}^{(0)}
  + \frac{5}{3} \alpha_{i,j}^{3} b_{3/2}^{(1)}\nonumber\\
  &\approx&-1.190494,\nonumber\\
C_{j,i}^\text{dir} &=& -b_{3/2}^{(1)}
  + \frac{5}{2} \alpha_{i,j} b_{3/2}^{(0)}
  - \frac{3}{2} \alpha_{i,j}^{2} b_{3/2}^{(1)}\nonumber\\
  &\approx&1.688311,\nonumber\\
C_{i,j}^\text{ind} &=& 0,\nonumber\\
C_{j,i}^\text{ind} &=& 1,
\end{eqnarray}
for a 2/1 resonance between $i$ and $j$ (planets pairs 1,3 and 2,4),
\begin{eqnarray}
C_{i,j}^\text{dir} &=& \frac{4}{5} \alpha_{i,j}^{-2} b_{3/2}^{(1)}
  - \frac{6}{5} \alpha_{i,j}^{-1} b_{3/2}^{(0)}
  + \frac{31}{30} b_{3/2}^{(1)}
  - \frac{11}{10} \alpha_{i,j} b_{3/2}^{(0)}\nonumber\\
  &&+ \frac{23}{15} \alpha_{i,j}^{2} b_{3/2}^{(1)}
  - \frac{16}{5} \alpha_{i,j}^{3} b_{3/2}^{(0)}
  + \frac{32}{15} \alpha_{i,j}^{4} b_{3/2}^{(1)}\nonumber\\
  &\approx&-2.025223,\nonumber\\
C_{j,i}^\text{dir} &=& -\alpha_{i,j}^{-1} b_{3/2}^{(1)}
  + \frac{3}{2} b_{3/2}^{(0)}
  - \frac{3}{2} \alpha_{i,j} b_{3/2}^{(1)}
  + 3 \alpha_{i,j}^{2} b_{3/2}^{(0)}
  - 2 \alpha_{i,j}^{3} b_{3/2}^{(1)}\nonumber\\
  &\approx&2.484005,\nonumber\\
C_{i,j}^\text{ind} &=& 0,\nonumber\\
C_{j,i}^\text{ind} &=& 0,
\end{eqnarray}
for a 3/2 resonance (planets pair 2,3),
and
\begin{eqnarray}
C_{i,j}^\text{dir} &=& \frac{32}{35} \alpha_{i,j}^{-3} b_{3/2}^{(1)}
  - \frac{48}{35} \alpha_{i,j}^{-2} b_{3/2}^{(0)}
  + \frac{36}{35} \alpha_{i,j}^{-1} b_{3/2}^{(1)}
  - \frac{36}{35} b_{3/2}^{(0)}\nonumber\\
  &&+ \frac{17}{14} \alpha_{i,j} b_{3/2}^{(1)}
  - \frac{93}{70} \alpha_{i,j}^{2} b_{3/2}^{(0)}
  + \frac{64}{35} \alpha_{i,j}^{3} b_{3/2}^{(1)}\nonumber\\
  &&- \frac{132}{35} \alpha_{i,j}^{4} b_{3/2}^{(0)}
  + \frac{88}{35} \alpha_{i,j}^{5} b_{3/2}^{(1)}\nonumber\\
  &\approx&-2.840432,\nonumber\\
C_{j,i}^\text{dir} &=& -\frac{16}{15} \alpha_{i,j}^{-2} b_{3/2}^{(1)}
  + \frac{8}{5} \alpha_{i,j}^{-1} b_{3/2}^{(0)}
  - \frac{19}{15} b_{3/2}^{(1)}
  + \frac{13}{10} \alpha_{i,j} b_{3/2}^{(0)}\nonumber\\
  &&- \frac{53}{30} \alpha_{i,j}^{2} b_{3/2}^{(1)}
  + \frac{18}{5} \alpha_{i,j}^{3} b_{3/2}^{(0)}
  - \frac{12}{5} \alpha_{i,j}^{4} b_{3/2}^{(1)}\nonumber\\
  &\approx&3.283257,\nonumber\\
C_{i,j}^\text{ind} &=& 0,\nonumber\\
C_{j,i}^\text{ind} &=& 0,
\end{eqnarray}
for a 4/3 resonance (planets pairs 1,2 and 3,4).
The coefficients $b_{3/2}^{(0)}$ and $b_{3/2}^{(1)}$
are the Laplace coefficients \citep[e.g.,][]{laskar_stability_1995},
evaluated at $\alpha_{i,j}$.

\section{Fixed points at first order}
\label{sec:fixedpoints}

In this appendix I describe in more detail
how to determine the position of the fixed
points of the averaged problem, at first order in eccentricities.
In this case, the Hamiltonian takes the form
\begin{equation}
  \H = \H_0(\epsilon, \Delta L_i)
    + \sum_{i,j\neq i} C_{i,j} \sqrt{D_i} \cos(\sigma_i + \vec{p}_{i,j}.\vec\phi),
\end{equation}
where $\vec{p}_{j,i} = \vec{p}_{i,j}$ are vectors of $n-2$ known integer coefficients
(which depend on the considered resonances),
and $\vec\phi$ is the vector of $\phi_i$ ($i \leq n-2$).
The fixed points of the averaged problem are solutions
of the following set of equations (see Eq.(\ref{eq:Hamjaceqs}))
\begin{eqnarray}
  0 &=& \frac{\partial\H}{\partial D_i} = \frac{\partial\H_0}{\partial\epsilon}
  + \frac{1}{2\sqrt{D_i}} \sum_{j\neq i} C_{i,j} \cos(\sigma_i + \vec{p}_{i,j}.\vec\phi)\\
  0 &=& -\frac{\partial\H}{\partial\sigma_i} =
  \sqrt{D_i} \sum_{j\neq i} C_{i,j} \sin(\sigma_i + \vec{p}_{i,j}.\vec\phi)\\
  0 &=& \frac{\partial\H}{\partial\Delta L_i} = \frac{\partial\H_0}{\partial\Delta L_i}\\
  0 &=& -\frac{\partial\H}{\partial\vec{\phi}} =
  \sum_{i,j\neq i} \vec{p}_{i,j} C_{i,j}  \sqrt{D_i} \sin(\sigma_i + \vec{p}_{i,j}.\vec\phi),
\end{eqnarray}
from which I deduce
\begin{eqnarray}
  \label{eq:solDsig}
  \sqrt{D_i} \expo{\im \sigma_i} &=& - \frac{1}{2}\left(\frac{\partial\H_0}{\partial\epsilon}\right)^{-1}
    \sum_{j\neq i} C_{i,j} \expo{-\im \vec{p}_{i,j}.\vec\phi},\\
  \label{eq:solphi}
  0 &=& \sum_{i,j\neq i,r\neq i} \vec{p}_{i,j} C_{i,j} C_{i,r} \sin\left((\vec{p}_{i,j}-\vec{p}_{i,r}).\vec\phi\right).
\end{eqnarray}
The parameter $\delta/\Gamma$ only appears in these equations
through the value of $\frac{\partial\H_0}{\partial\epsilon}$.
Therefore, at first order in eccentricities, all the angles, as well as the eccentricity ratios are independent of $\delta/\Gamma$.
The value of this parameter only changes a factor common to all eccentricities
(see Eq.~(\ref{eq:solDsig})).
Equation~(\ref{eq:solphi}) provides a set of $n-2$ equations on the $n-2$ angles $\phi_i$.
For instance, for a 3:4:6:8 resonant chain such as \object{Kepler-223},
I obtain (see Eqs.~(\ref{eq:hamdLaDk223}) and (\ref{eq:solphi}))
\begin{eqnarray}
 \label{eq:solphik223}
0 &=& 2C_{1,2}C_{1,3}\sin(2\phi_1)\nonumber\\
  && +3C_{2,1}C_{2,3}\sin(3\phi_1) + 3C_{2,1}C_{2,4}\sin(3\phi_1+\phi_2)\nonumber\\
  && + C_{3,1}C_{3,2}\sin(\phi_1) + C_{3,1}C_{3,4}\sin(\phi_1+2\phi_2),\nonumber\\
0 &=& C_{2,1}C_{2,4}\sin(3\phi_1+\phi_2) + C_{2,3}C_{2,4}\sin(\phi_2)\nonumber\\
  && + 2C_{3,1}C_{3,4}\sin(\phi_1+2\phi_2) + 2C_{3,2}C_{3,4}\sin(2\phi_2)\nonumber\\
  && + C_{4,2}C_{4,3}\sin(\phi_2).
\end{eqnarray}
There are trivial solutions at $0$ and $\pi$,
but other (asymmetric) solutions might exist.
The existence of asymmetric solutions is due to
the influence of first-order resonances between non-consecutive pairs.
In the case where only consecutive pairs are involved in resonances,
Eq.~(\ref{eq:solphi}) simplifies, and it can be shown that only symmetric solutions exist.
The number of solutions of a highly non-linear set of equations such as
Eq.~(\ref{eq:solphik223}) is not easily predicted.
Moreover, among those solutions, some correspond to elliptical (stable) fixed points,
and the others to hyperbolic (unstable) fixed points.
In the case of \object{Kepler-223},
I solve for the position of fixed points numerically,
and only consider the stable fixed points.
I find six possible stable solutions.

Once a solution is found for $\phi_i$ ($i \leq n-2$), the angles $\sigma_i$,
and the eccentricity ratios can easily be deduced from Eq.~(\ref{eq:solDsig})
\begin{equation}
  \label{eq:solesig}
   e_i \expo{\im \sigma_i} \approx e_0 \sum_{j\neq i} \frac{m_j}{m_0} C'_{i,j} \expo{-\im \vec{p}_{i,j}.\vec\phi},\\
\end{equation}
with
\begin{eqnarray}
  C'_{i,j} &=&
   - \frac{a_{n,0}^{3/2}}{a_{\max(i,j),0} \sqrt{a_{i,0}}} \left( C_{i,j}^\text{dir}
  - \frac{C_{i,j}^\text{ind}}{\sqrt{\alpha_{i,j}}}\right),\nonumber\\
  e_0 &=& - n_{n,0} \left(\frac{\partial\H_0}{\partial\epsilon}\right)^{-1}.
\end{eqnarray}

For a 3:4:6:8 resonant chain such as \object{Kepler-223}, I obtain
\begin{eqnarray}
   \frac{e_1}{e_0} \expo{\im (4\lambda_2 - 3\lambda_1 - \varpi_1)} &\approx&
      6.2526 \frac{m_2}{m_0}
     + 1.9999 \frac{m_3}{m_0} \expo{-\im 2\phi_1},\nonumber\\
   \frac{e_2}{e_0} \expo{\im (3\lambda_3 - 2\lambda_2 - \varpi_2)} &\approx&
     - 6.5665 \frac{m_1}{m_0} \expo{\im 3\phi_1}
     +3.0911 \frac{m_3}{m_0}\nonumber\\
     &&+1.4999 \frac{m_4}{m_0}  \expo{-\im \phi_2},\nonumber\\
   \frac{e_3}{e_0} \expo{\im (4 \lambda_4 - 3 \lambda_3 - \varpi_3)} &\approx&
     -0.5712 \frac{m_1}{m_0} \expo{\im (\phi_1+2\phi_2)}
     - 3.3120 \frac{m_2}{m_0} \expo{\im 2\phi_2}\nonumber\\
     &&+3.1263 \frac{m_4}{m_0} ,\nonumber\\
   \frac{e_4}{e_0} \expo{\im (4 \lambda_4 - 3 \lambda_3 - \varpi_4)} &\approx&
     -0.4284 \frac{m_2}{m_0} \expo{\im \phi_2}
     - 3.2833 \frac{m_3}{m_0}.
\end{eqnarray}

\end{document}